BC10991

# Prediction of High Transition Temperatures in Ceramic Superconductors

J. C. Phillips

Dept. of Physics and Astronomy, Rutgers University, Piscataway, N. J., 08854-8019

## Abstract

The prediction of transition temperatures can be regarded in several ways, either as an exacting test of theory, or as a tool for identifying theoretical rules for defining new homology models. Popular "first principle" methods for predicting transition temperatures in conventional crystalline superconductors have failed for cuprate HTSC, as have parameterized models based on $CuO_2$ planes (with or without apical oxygen). Following a path suggested by Bayesian probability, we find that the glassy, self-organized dopant network percolative model is so successful that it defines a new homology class appropriate to ceramic superconductors. The reasons for this success are discussed, and a critical comparison is made with previous theories. The predictions are successful for all ceramics, including new non-cuprates based on FeAs in place of $CuO_2$.

**1. Introduction**

The prediction of transition temperatures $T_c$ is rightly considered to be one of the most difficult problems in theoretical physics. Here one should distinguish between a true (or bare) prediction (the value of $T_c$ is not known when the prediction is made), and a postdated calculation carried out according to certain rules after $T_c$ has been measured experimentally. If the rules (sometimes called "first principles") have been established for similar materials, and they are faithfully applied to the new case, then the validity of the rules can be tested by the success of the prediction. If the number of example materials where the rules have been applied previously is large compared to the number of adjustable parameters, then the prediction can be said to be based on homology, namely the supposed microscopic similarity of the phase transition of the new material to those previously studied.

Perhaps the best-known example of a "bare" prediction of a transition temperature was [1] for superfluidity of $^3$He. The predicted value of $T_c$ was 100 mK, whereas the measured value [2] of $T_c$ was 3.6 mK, so the predicted value was too large by a factor of ~30. The most successful predated prediction of $T_c$ for superconductors based on homology was for the high-pressure phase of Si [3], where the experimental maximum $T_c$ is 8.2K, and estimation of $T_c$ using the most popular rules gave $T_c$ = 5K. A number of postdated predictions of superconductive $T_c$'s have been made based on homology models and "first principles" rules. The rules depend mainly on the Fermi-surface average of the electron-phonon coupling constant $\lambda$ and the phonon frequency squared ($\lambda<\omega^2>$). The most popular of these have been the prediction of $T_c$ in $MgB_2$ based on homologies with simple metals such as Al, where the predicted and experimental values are both about 40K [4]. Equally impressive has been the prediction of $T_c$ in 5% B-doped CVD diamond, where the predicted and experimental values are $T_c$ = 22K and $T_c$ = 11K, respectively [5]. The agreement here between theory and experiment is better than it appears to be, as some of the B may have formed electrically inactive dimers. This example illustrates the strengths and limitations of homology arguments based on first principles, as the material properties of B-doped CVD diamond are a far cry from those of single-crystal Al or $MgB_2$.

Other examples of $T_c$ predictions are the proton order-disorder transition in high-pressure ice, where the predicted and experimental values are 98K and 72K, respectively (KOH doping) [6], and strained thin-film ferroelectrics, where large shifts in the Curie temperature $T_c$ are predicted, and there is good agreement between theory and experiment after correcting for domain effects [7]. In magnetic materials the values of $T_c$ calculated by self-consistent fields with frozen magnons are in generally good agreement with experiment within about 20%, even for complex ternary alloys with mixed ferro- and antiferro-magnetic interactions [8]. All of these successful calculations have occurred in well-ordered crystals.

**2. Exotic Superexchange and Traditional Electron-Phonon Interactions**

Much of the theoretical literature over the last two decades has been devoted to discussing non-phonon models of HTSC. Did Bednorz and Mueller suppose, when they discovered HTSC in the cuprates, that they were also opening what would turn out to be a Pandora's box for theorists? Certainly by looking for superconductivity in a family of materials known not for metallic, but for antiferromagnetic properties, they were already behaving as contrarians, who worked evenings on this project, as they felt that this research direction would not meet with the approval of their immediate superiors. Had they consulted me, I would have asked them why they expected to find superconductivity in these pseudoperovskites. Their explanation – that the strongest electron-phonon interactions are found in the best ferroelectrics (like $BaTiO_3$, which has the closely related perovskite structure) would not have satisfied me. The perovskites exhibit many displacive distortions, and a would-be metal should be rendered insulating by Jahn-Teller distortions.

In fact, my general argument was correct for almost all such materials – almost all, but with a few exceptions! As I later learned, the crystal chemist who discovered $La_2CuO_4$ had made a prescient remark (references for this and many other older papers can be found in my book [9]). He noted that $La_2CuO_4$ was the only cuprate known at that time with an undistorted tetragonal structure. In all other cuprates the lattice planes had buckled due to strong Jahn-Teller distortions, and these same distortions would suppress metallic conductivity and of course superconductivity. No one could blame theory (or me) for not predicting the rigidity of $CuO_2$ planes, and I was off the hook for the moment, but there were much bigger challenges yet to come.

These new challenges emerged in a way that took me even more by surprise than the Bednorz-Mueller discovery. It had been known for decades that superconductivity and magnetism are incompatible, because electron-spin scattering breaks Cooper pairs. There are elaborate many-body ways of deriving this result (based on four-particle scattering diagrams called plaquettes), but the simplest way is to invoke Helmholtz' theorem, which states that any vector field can be decomposed into complementary parts, solenoidal (magnetic) and irrotational (extensive, in other words, superconductive). This is, in fact, a "poor man's" way of deriving the Meissner effect.

Soon phase diagrams appeared that showed that as the doping level increased, magnetism faded and superconductivity appeared, just as one would have expected. However, even before phase diagrams became available, theorists had jumped in with explanations, and the first explanation was not the most natural one, but the most surprising one, even more surprising than the discovery itself. Anderson proposed that HTSC was caused not by phonons, but by superexchange between spins (!), a subject that he had discussed in 1950 for insulators. It seems that he had long hoped to find superexchange in metals, and it appeared to him that HTSC were just what he had been looking for. He termed his new mechanism "RVB", and it soon became by far his most cited paper (~ 5000 citations so far). Anderson is still advocating RVB [10] and that HTSC was caused not by phonons, but by superexchange between spins (!), and he has persuaded many distinguished theorists to join him [11]. His example has stimulated many frivolous theories.

Of course, not everyone was in love with superexchange. Sir Neville Mott had always championed a common-sense approach (today known more popularly by the computer scientists' acronym KISS), and initially he suggested that HTSC could be caused by a mixture of electron and spin interactions. However, he soon abandoned spin altogether, and turned to "bipolarons", by which he meant Cooper pairs formed by very strong and very localized electron-phonon interactions [12,13]. However, while there is no doubt that very localized electron-phonon interactions exist at dopants (such as interstitial O) in these materials, it is not easy, especially with such large unit cells and so many normal modes, to distinguish experimentally such dopant interactions from host Jahn-Teller distortions. Indeed, even in the old intermetallic superconductors, with $T_c$ ~ 20+K, it was often Jahn-Teller distortions (or other lattice instabilities) that had ultimately limited $T_c$ [9], while magnetic interactions, even with small amounts of magnetic impurities, quickly destroyed superconductivity. T. D. Lee also suggested that HTSC might be more like Bose-Einstein condensation, and there are still papers on HTSC using the B-E approach. In my opinion, this approach is insufficiently material-specific, and does not identify the key aspects that make the cuprates special.

From the point of view of materials science, what was special about the cuprates was (and still is) that they are at the cutting edge (sometimes called the bleeding edge) of new materials: not only are they complex chemically, but oxides *per se* previously enjoyed a very poor reputation among crystallographers: the samples were often oxygen deficient, and the sample quality was often so poor that X-ray structural determinations did not meet the standards required for publication in archival diffraction journals. It seemed to me [14] that these problems suggested that these materials should be regarded as Mechanically Marginally Stable (MMS). MMS is apparently unapproachable theoretically; the reason is that too little is known about interatomic forces, and even small errors in these will cause the dynamical matrix to produce negative values of $\omega_\alpha^2(q)$ for some q and some mode α. The current state of the art (2009) is that MMS in ferroelastic and ferroelectric $BaTiO_3$ (but not $La_2CuO_4$) has been solved by brute-force first-principles pseudopotential calculations. However, even the host compound $La_2CuO_4$ is a very long way from $La_{2-x}Sr_xCuO_4$ (the dopants are always ignored, as they are a large problem for brute-force approaches, but as we shall see, there are other ways to deal with them).

When $T_c$ is calculated for HTSC using crystalline metallic ($\lambda<\omega^2>$ or Eliashberg) rules, not only are the predicted values too low, but also the chemical trends are wrong (with some fudging of the adjustable Coulomb repulsion parameter $\mu^*$, $T_c$ in $La_{2-x}Sr_xCuO_4$ may be brought close to experiment (x = 0.15, $T_c$ = 38K), but the same value of $\mu^*$ in $YBa_2Cu_3O_{7-x}$ gives $T_c$ < 1K, a failure compared to experiment (x = 0.1, $T_c$ = 90K) [15]. The feature that distinguishes ceramic HTSC from metals is their strong disorder, as reflected in complex patterns of nanodomains on a length scale ~ 3 nm [16,17]. In the presence of strong disorder, superconductivity may become molecular in character [12,13], in which case $T_c$ could be as high as 380K [18]. Unfortunately, this upper bound is unlikely to be reached, because in the presence of strong electron-phonon interactions the metallic band at $E_F$ will be split by a Jahn-Teller effect (for example, monovalent dopants could form electrically inactive dimers); worse still, the entire compound may not be dopable, or may phase-separate.

3. **Chemical Factors: Size, Apical Oxygens, Electronegativity and Valence**

A different and more global approach to strong disorder relies on traditional analysis of chemical factors (valid in crystals, molecules and glasses) [19] to analyze trends in $T_c$. Some have argued that cuprate superconductivity must be localized in the $CuO_2$ planes [20], but if this were the case, the maximum $T_c$'s in each material family, $T_c^{max}$, would be nearly constant, much like the planar lattice constant. This is far from being the case. As one can easily see from structural systematics [19], the central problem in ceramic superconductors lies in making a ceramic conducting by heavy doping. Light doping (as in semiconductors) is not enough, because as the ceramic dielectric constant is small, the dopant orbitals are also small, and these will not overlap unless the dopant level is high. Maximum doping levels in rigid semiconductors are typically < 1%, so something special is required to support the high doping levels found in ceramic superconductors and avoid phase separation.

In cuprates that special factor is the exceptional rigidity of $CuO_2$ planes (see Sec. 2), which allows the intervening insulating layers to be soft and flexible. Because core repulsive potentials are "hard", it is then relatively easy to dope the soft insulating layers. Note that normally one thinks of metals as soft, and ionic insulators as hard. The cuprates and a few other layered ceramics reverse the normal ordering, and this is what makes it possible for them to be HTSC. There are no mysterious superexchange interactions between spins; most of the antiferromagnetic regions have disappeared by the time the doping is large enough for the materials to be metallic. The pseudogaps are still present, and they play a necessary role, but whether or not they are caused by magnetic interactions (unlikely) or charge density waves (probably) is incidental. The highest $T_c$, reached at optimal doping in a given alloy, will depend on the dopant configuration relative to the regions occupied by the pseudogaps. That in turn will depend on the relative sizes of the ions, a packing (actually, a dynamical packing) question that is well beyond the reach of present and probably any future theory. As we will see in the next section, with a sufficiently large data base this difficulty can be handled by advanced statistical methods.

Kamimura recognized that the $CuO_2$ planes alone could not explain HTSC, no matter how exotic the interactions within them, and he suggested that apical oxygens could play an important role [21]; several succeeding "apical oxygen" papers also proved quite popular [22,23]. There have always been strong indications that apical oxygens are coupled to interlayer dopants (especially interstitial oxygen, which is close to apical oxygens and can form dynamical vibrational bands with them) based on chemical trends between apical oxygen bond lengths, ordering and $T_c$ [24], and we shall see further evidence of such correlations later. However, specific apical oxygen models always introduced atomic orbitals and their overlaps, which generate many parameters to fit one observable, a situation that always leads to "excellent" agreement between "predicted" and experimental values. In fact, the most popular model [23] is based on the difference between the Madelung potentials at the planar and apical oxygen sites, calculated with an artistic point ion model where most of the many charges in the unit cell are freely "assigned" (for instance, although O is usually 2-, in some cases it is "assigned" a charge of 1+, and many liberties are taken with "assigned" cationic charges). A parameterized discussion of correlations between apical oxygen bond lengths $d_a$ and $T_c$ [25] gave plausible trends, but it covered only a few cases with no predictions of $T_c$, showing the limitations of $d_a$ as a configuration coordinate. More generally, one should not use extensively parameterized atomic models ($CuO_2$ planes + apical oxygens) to discuss differences between pseudogapped regions and superconductive regions, as this difference is almost surely due to different dopant distributions [26].

When one considers seven different structural factors, including apical oxygen bond lengths, and correlates them with $T_c$, one finds that the obvious ionic variable, the average over *all* host atoms of the difference between cation and anion electronegativities ($<\Delta X>$), gives the best fit to $T_c$ [27]. (This is the ionic analogue of the covalent molecular model [14].) Bearing in mind that different dynamical packings can be used to maximize $T_c$, we can now test the validity of $<\Delta X>$ as a coupling parameter by plotting $T_c^{max}$ ($<\Delta X>$). The results are quite disappointing, as they yield a scatter-shot plot [28], which means that the traditional chemical approach is too simple to explain HTSC (the reason is that to produce a strong electron-phonon interaction, we need a soft lattice, and soft lattices will be found near a covalent-ionic metallic triple point, which will not depend on $<\Delta X>$ alone). It is just at this point, when nothing seems to work, that we can find the answer, using only the elements discussed so far.

## 4. Successful Prediction of $T_c^{max}$

We begin by realizing that volume factors, although they are not known in detail, can be included implicitly in the analysis by focusing initially not on $T_c(Y)$, but rather on the largest transition temperatures $T_c^{max}$ (Y), where Y is any other chemical factor. In other words, if the volume factor has already been optimized, then the material in question will have $T_c = T_c^{max}(Y)$. For those familiar with Bayesian probability, this approach is readily recognized: Bayesian probability interprets the concept of probability as a contingent "measure of a state of knowledge", and not as a frequency in orthodox statistics (further details can be found from your browser). Bayesian methods can be extremely effective, but the Bayesian filter requires a large data base. In what follows, the basic Bayesian conclusions are tested by employing the full ceramic HTSC data base, which has now grown quite large, including ceramics based on metallic layers other than $CuO_2$ planes and apical ions other than oxygen.

At this point the only chemical configuration coordinate remaining is valence, but this coordinate can compete with structural coordination numbers [14]. The latter are much the same for most cuprates, and are unlikely candidates for Y. Moreover, decades of research on network glasses have shown (particularly when comparing silicate and chalcogenide alloy glasses) that the marginal mechanical stability of such densely packed glasses (which permits them to avoid crystallization) depends primarily on the number of Pauling resonating valence bonds R (for example, in NaCl there is one Pauling resonating valence bond/ion, although the coordination numbers are six) [29]. The value of R for Cu in the cuprates is 2. There are similarly obvious rules for R for other elements (including those with mixed valence, Tl, +1,+3 and Bi, +3,+5) [30]). The conclusion (also taking into account many results for network glasses) is that while R is a good coordinate for ionic and covalent molecules, it is a very accurate configuration coordinate for strongly disordered networks, even in the presence of lone pair interactions [28].

**5. MMS and Coarse Graining**

The question remains of how valence or R is to be averaged over the many atoms in the unit cell: should the rigid metallic planes be weighted more heavily, or the soft layers between them, where the dopants are? Again decades of research on network glasses have shown that the best weighting is the simplest, that is, equal weighting (<R>) for all atoms provides the best description, because the overall network is Mechanically Marginally Stable (MMS). As discussed above, shell-model calculations [31] of phonon spectra in HTSC are generally unable to predict very soft modes, especially those associated with dopants. Quite revealing is the fact that the fitted effective charges for "soft layer" rare earth cations in shell models of LSCO and YBCO were inexplicably *negative* [32]. Neutron scattering has successfully identified Jahn-Teller distortions associated with optical modes, and these set the overall scale [30] for $T_c^{max}$, but it is unable to resolve very soft modes [33].

Although MMS is concealed from experiment and theory in HTSC, the simple and universal assumption of equal weighting of all atoms is fully effective. In retrospect it is easy to see why equal weighting works so well. HTSC exhibit nanodomains spanning ~ 10 unit cells [16,17], so the soft modes implied by MMS are spread not only over a single unit cell, but in fact over many unit cells (coarse graining), all of which are sufficiently disordered by dopants and pseudogap instabilities that equal weighting becomes almost as accurate as in fully homogeneous statistical systems.

Guided by these considerations, which are general, specific, simple, and free of adjustable parameters, we plotted $T_c^{max}$ (<R>), with the results shown in Fig. 1(a). It is important to understand how the predictive dotted line was drawn. First, note that all the HTSC with $R \geq 2$ are well-known, and in fact were discovered chronologically with decreasing <R>, starting with $Ba_{1-x}K_xBiO_3$ (a cubic perovskite, with a phase diagram quite different from the cuprates, and only hints of an emerging nanodomain structure [30], but it still fits nicely on the smooth dotted line). Secondly, when the predictive dotted line was drawn initially, only one point was known for <R> < 2, that for NCCOC ($Na_xCa_{2-x-y}CuO_2Cl_2$) with an apical Cl (not O) and $T_c^{max}$ ~ 40K. This point is approximately the mirror image of LSCCO about <R> = 2, so it seems natural to draw the predictive line as shown. Later $Ba_2Ca_2Cu_3O_6F_2$ (apical F) was reported to have $T_c$ = 55K, which is comfortably below $T_c^{max}$ = 78K on the predicted line, so this point was added to the published Figure [30] as its first "predictive" success. Finally improved sample preparation [34] gave $T_c$ = 76K, reducing the discrepancy between the predicted and experimental value of $T_c$ from 23K to 2K. This is a very convincing second (and predated) predictive success, especially considering that neither NCCOC nor BCCOF contain apical O, which means that the homology is unexpectedly general.

The analysis given here is based on minimal logical and statistical considerations, but the results were actually derived from a physical model that the author first proposed 20 years ago [35]. The zigzag self-organized percolative model has many attractive features: for instance, the energy scale for $T_c^{max}$, which is merely set empirically in Fig. 1, is apparently set for $\langle R \rangle \geq 2$ by the Jahn-Teller shift in the (100) LO phonon energy (there are no data for $\langle R \rangle < 2$), as shown by Fig. 1(b) in [30], which suggests a close dopant-mediated relation between the soft acoustic and optic modes. Substantial evidence shows that percolative filaments are formed at high temperatures and account for many features of transport up to at least 300K [30]. At present there exists no alternative model that can predict $T_c^{max}$ in HTSC.

## 6. Cuprate-Like Superconductivity

There are several new marginally stable families of layered crystals that exhibit many similarities to the cuprates: there are many atoms per unit cell, with displacive lattice instabilities, vicinal antiferromagnetic phases, etc., including ionic $Li_xZrNCl$ [36] and the rapidly growing covalent superfamily based on FeAs, such as $LaFeAsO_{1-x}F_x$ [37]. Are these similarities accidental, or can the percolative cuprate ionic superfamily model explain HTSC in these non-cuprate layered materials *with no additional assumptions*? In fact, the percolative model *easily* explains the similarities, by bringing these new materials into the general framework of self-organized networks. This has already been done for ionic $Li_x(Zr,Hf)NCl$ ($T_c \sim$ 15K-25K)[38], so now a similar discussion is given here for the covalent $LaFeAsO_{1-x}F_x$ superfamily ($T_c \sim$ 26K-43K) [37], which is much larger and the subject of hundreds of recent studies.

Because these materials are mechanically only marginally stable, they are strongly disordered when doped, and are generally far from optimized with respect to HTSC. Marginal lattice stability determines the overall scale for $T_c$, as the phonon energy shift measured by neutron scattering associated with Jahn-Teller doubling of the unit cell of LO phonons correlates linearly with $T_c^{max}$ in the cuprates [30]; of course, spin and antiferromagnetic exchange show no such scaling, and are irrelevant. The least upper bound for $T_c$, called $T_c^{max}$, has a strongly percolative character, as it peaks exponentially at $\langle R \rangle = 2$, where $\langle R \rangle$ is the average valence number of all the atoms [28,30].

The master function $T_c^{max}(\langle R \rangle)$ is shown in Fig. 2 for the cuprates; the peak at $\langle R \rangle = 2$ has a cut-off exponential character. The point for ionic $Li_xHfNCl$ was discussed previously [38], and we now discuss the points for the LaFeAsO$_{1-x}$F$_x$ family (T$_c$ ~ 26K-43K). Here R(Fe) = 2, just as for Cu, because Fe is in a 2+ valence state [39] in a virtual crystal model. (This band model also shows the beginnings of self-organization, in that the average height h of As is found to shift with x [40]; had the calculation been carried out with a large supercell centered on a F dopant, h(As) would have varied with distance from F. This virtual crystal relaxation effect alone shows that N(E$_F$) varies slowly and smoothly with doping (which means that the apparent Fermi planar line becomes very broad near a critical point), and hence effective medium models cannot explain HTSC. However, from this one should not conclude that electron-phonon interactions do not cause HTSC, as these interactions (not superexchange!) do set the overall energy scale [30] through Jahn-Teller distortions on and near percolative paths, and these distortions are especially large near dopants.

When one calculates <R> for undoped LaFeAsO, one obtains <R> = 2.5, which places LaFeAsO$_{1-x}$F$_x$ (T$_c$ = 26K) very close to Li$_x$ZrNCl on the master curve of Fig. 2, so that the theory appears to succeed effortlessly. However, T$_c$ is maximized at 43K for pressures near 4 GPA [41], an increase of 60%, which is much more than is seen in cuprates, and exceeds the upper bound of the master curve. Does this falsify the theory? No, because the equilibrium cuprate master function T$_c^{max}$ (<R>) remains valid for the Li$_x$ZrNCl and LaFeAsO$_{1-x}$F$_x$ families, but the pressure dependence in the latter family is larger than in the cuprates, as Fe-As bonding is more covalent than largely ionic Cu-O bonding [41], due to the larger Pauling electronegativity X differences (X(Fe) = 1.8, X(As) = 2.0, X(Cu) = 1.9, X(O) = 3.5)) in the cuprates. Covalent bonding also seems to limit the range of <R> for which stable crystals can form; thus the smallest value of <R> for the FeAs family seems to be near 2.3 in (Ba$_{0.55}$K$_{0.45}$)Fe$_2$As$_2$ [42].

It is striking that this covalent FeAs stability range of 2.3 < <R> < 2.5 is very similar to the range 2.25 < <R> < 2.52 for stress-free covalent glasses previously shown [28,30] in Fig. 2. (Again this is a rather remarkably successful predated prediction.) Thus the covalent instabilities of the host lattice, especially the reduced rigidity of the FeAs plane relative to the CuO$_2$ plane, are the factor that shifts the value of <R> at which T$_c^{max}$ peaks from <R>$_{max}$ = 2.0 in the ionic cuprates to <R>$_{max}$ = 2.5 in the LaFeAsO$_{1-x}$F$_x$ family. In both cases

$$\langle R \rangle_{max} = R(\text{metallic plane}) \qquad (1)$$

so that the average connectivity of the undoped insulating plane is matched to that of the metallic plane. Apparently this hidden topological layer symmetry (1) determines the maximum electron-phonon interactions in marginally stable layered pseudoperovskites.

## 7. Surfaces and Interfaces

The percolative master function $T_c^{max}$ ($\langle R \rangle$) is determined from bulk data on layered crystals, so one can ask whether or not this function can explain trends in $T_c^{max}$ at surfaces and layer interfaces. Determining $\langle R \rangle$ at surfaces and interfaces is much more difficult than in the bulk, where it is natural to average R over all atoms, as the self-organized structure is marginally stable overall, and soft modes that are critically bound to percolative superconductive paths should have long wave lengths, as $T_c^{max}$ is still small compared to the melting temperature. However, in layered thin crystalline films with epitaxial interfaces or at doped free surfaces, similar percolative behavior is expected. This turns out to be the case for the $La_2CuO_4$ - overdoped $(La,Sr)_2CuO_4$ interface, where $T_c \sim 50K$, after enhancement by exposure to ozone from ~ 30 K [43]. The ozone enhancement is readily explained by the addition of oxygen dopants, absorbed by $La_2CuO_4$ to give $La_2CuO_{4+\delta}$ with $\delta \sim 0.15$. However, the maximum $T_c$ obtainable in this way is ~ 30K for both layers separately, apparently producing a mystery.

Let us look at this mystery with the master curve. $\langle R \rangle$ = 16/7 in LCO = 2.28, = (16 - x)/7 in $La_{(2-x)}Sr_xCuO_4$, so with x (or $\delta$) = 0.15, $\langle R \rangle$ = 2.26, and with x = 0.45, $\langle R \rangle$ = 2.22. This would give a decrease in $\langle R \rangle$ between x = 0.15 and x = 0.45 of 0.04, and so we get something like $T_c$ = 35K + (0.04/0.28)[150-35]K = 50K. Of course, this is just a plausible guess at the interfacial structure, but the master function has given the trend correctly, not only qualitatively, but even semi-quantitatively (something no other theory has been able to do: virtual crystal theories predict $T_c$ < 1K, from which it has often been erroneously concluded that electron – phonon interactions do not cause HTSC!).

Now we turn to a much more difficult problem, for which the data base is small, but still robust: a surface monolayer of $A_xWO_3$, where A is an alkali metal (Na [44] or Cs[45]). While bulk

Na$_x$WO$_3$ exhibits superconductivity only near 1K, here for Na superconductivity appears around 100K; for Cs there are two phase transitions, a bulk one with lower T$_c$ at higher doping, and a re-entrant percolative one with higher T$_c$ at lower doping. Moreover, Na- and Li- (but not K-) doped surfaces of nanoclusters of WO$_3$ embedded in a variety of nanoporous hosts (carbon inverse opal, carbon nanotube paper, or platinum sponge) show diamagnetic anomalies with an onset T of 130K [46]. Note that WO$_3$ (with its simpler unit cell, subject only to Jahn-Teller distortions) itself is nonmagnetic, as is another HTSC (BKBO, (Ba,K)BiO$_3$).

These data can be combined with the master function T$_c^{max}$(<R>) to construct a model of percolative self-organization at surfaces. In bulk WO$_3$ the valence of W is 6, and <R> = 3.0, far to the right on the master function T$_c^{max}$(<R>), where T$_c^{max}$(3) < 5K. Near the surface the valence of W could be 2 (just as with Cu in the cuprates, and Fe in the FeAs compounds). To explain T$_c$ ~ 130K one must assume <R> = 2. A percolative W$_s$O surface chain then has <R> = 2. These surface chains are entropically broken into stress-relieving fragments. Intercalated Li or Na ions connect the chain fragments, thereby increasing their conductivity and their screening of internal ionic fields, just as in the bulk percolative model. The embedded clusters are not connected, so the result is "localized non-percolative superconductivity", still with <R> = 2 [47]. In the free surface case [44,45] thermal fluctuations disrupt superconductivity above 100 K. Both of these <R> = 2 points are shown in Fig. 2. Considering the rapid progress in nanoscience, it may be possible to obtain similar pairs of percolative and cluster points for other HTSC.

Finally, self-organized percolation enables us to understand how cointercalation of organic molecules M with Li in ionic Li$_x$M$_y$HfNCl can uniformly enhance T$_c$(x) by up to 30%, over a wide range $0.15 \leq x \leq 0.50$, even though the average interlayer spacing d varies by as much as 30% [48]. The intercalated organic molecules reduce the dielectric screening of Li-centered electron-phonon interactions by other Li ions. Note that the nominal concentration x of the Li ions may refer only to patches that strongly diffract; the superconductive paths may pass through patches with a concentration x$_0$ different from x, explaining why T$_c$ is apparently constant over a wide range of x. (Only a few percolative paths are required to exclude Abrikosov vortices and produce HTSC; this is why T$_c$ often increases when the average density of states N(E$_F$) decreases (for instance, in the cuprates compared to the covalent FeAs family, or the ionic $Li_xM_yHfNCl$ family [38]). This is yet another example of the paradoxes generated by attempting to understand a self-organized percolative phase in terms of continuum concepts such as Bloch-wave Fermi surfaces, modulation doping or plasmon waves. Note that these points have often been made in earlier papers on the topological percolative model, long before any of these new materials were discovered.

8. **Mapping Percolative Paths**

For most of the last two decades experiment has provided largely circumstantial support for the zigzag percolative model, but recently more direct evidence has appeared. The model has two key elements: the dopants, often interstitial oxygen, whose positions are not easily determined, and the zigzag paths themselves. The zigzag paths are associated with strong electron-phonon interactions, which are especially strong for interlayer displacements. This has enabled the zig and the zag of the paths to be identified by a combination of time-resolved electron diffraction [49], for the c-axis zag component of the paths, and anisotropically strong kinks in quasi-particle dispersion observed by angle-resolved photoemission [50], for the ab planar zig component of the paths. It is difficult to understand these correlations unless the c axis zag component is actually topologically connected to the planar zig component, so the results are fully consistent with the zigzag model discussed in dozens of papers over the last 20 years. There remains one puzzling aspect: anisotropy is observed in both Bi2212 and Bi2223 by ARPES, but only in the former by electron diffraction. The extra layer in the latter would not appear to erase the planar anisotropy, but it is possible that Bi2223 has a high density of stacking faults. Depending on the geometry of the percolative paths, such faults could erase the observed anisotropy by scattering percolative carriers.

While we are on the subject of percolation vs. continuum models, it is important to realize that while the presence of self-organized percolative paths introduces exponential complexity into the Bohmian wave-packet basis states [51] used to form Cooper pairs, it in no way alters the nature of the attractive electron-phonon interactions responsible for forming the pairs. These two points have often been confused, and the failure of continuum plane wave basis states to describe cuprate HTSC has been used to argue that electron-phonon interactions do not cause HTSC [11]. Because the fraction of carriers involved in percolative paths is small, many negative indications of strong electron-phonon coupling have been found; for example, bulk phonon softening is so small as to be unobservable by neutron scattering at high energies [33]. Therefore it is gratifying that quasiparticle tunneling across a break junction *perpendicular* to the superconducting copper oxide planes showed 11 phonon features that match precisely with Raman spectra [52], decisively showing that HTSC is indeed caused by electron-phonon interactions along the c axis, as previously argued from dynamical relaxation experiments [53]. Also the long-expected isotope effect at the phonon kink has been observed by ARPES [54]. The fact that *all* the c-axis phonon bands appear in the break junction experiment, and not just those associated with the O buckling mode in the $CuO_2$ plane [50], strongly supports the present model of zigzag percolative paths in a marginally stable all-atom network. Finally, it is worth stressing that the phonon features that correlate by far the best with $T_c$ are exactly those of the atoms in the soft insulating layer (Figs. 11 and 12 of [55]), as assumed in the zigzag model, and not the atoms in the rigid $CuO_2$ layer, as is often wrongly assumed [50].

## 9. Conclusions

The present discussion shows that the original percolative model [35] continues to provide an excellent universal guide to the phenomenology of ceramic supercondcutors, as it quantitatively predicts superconductive $T_c$'s, even for non-Cuprate HTSC, starting only from the Cuprates, which no other theoretical model has done. The self-organized marginally stable percolative model unexpectedly transcends conventional chemistry (the materials covered include cuprates, tungstates, zirconates, ferrics, arsenides, bismates, oxides, halides, Li-, Na-, Cs-, O-, F-, Cl-, doped..., and the structures include bulk, interface and nanocluster surfaces); so far as I know, there is no other example in solid phase transitions where theory of any kind (continuum or otherwise) can predict transition temperatures with such universal success. The results make an overwhelming case for a new homology class of glassy doped superconductors, separate and distinct from continuum metallic superconductors.

As noted previously [30], the theory predicts $T_c^{max}$ with an accuracy of 10 K, compared to a melting point of order $10^3$ K, which is an accuracy of 1%. However, this success contains a subtle aspect, which is that because HTSC is concerned with marginally stable lattices, it is important to build this theoretical mechanism into the analysis from the outset (Sec. 4). This has been done by mapping $T_c^{max}$ (<Y>), and examining different choices of Y: only Y = R gives the resonant peak that is strongly suggestive of percolation.

The theoretical debates stimulated by HTSC have lasted for more than 20 years, and have often severely tried the patience of the physics community. Were they worth it? I believe they were and still are. The basic issue was, and still is, whether it is legitimate to treat a discrete, exponentially complex problem involving strongly disordered materials, at the cutting edge of materials science, with the same simplistic continuum methods (including polynomial Hamiltonian algebra) that students have learned work so well for toy models and some simple crystalline solids. The experimental evidence strongly suggests that this discrete, exponentially complex problem can be solved by focusing on its essential topological features, derived from the network structures of ceramic HTSC [29]. Topology is still something of an oddity among mathematical disciplines (it is very young, dating from ~ 1870). It is unfamiliar to most chemists and physicists, but it is ideally suited to treating complexity problems, including many far afield from science, such as economics [56,57] and difficult problems in biology, such as protein-protein interactions [58]. It also transcends the otherwise theoretically insuperable barriers of exponentially complex (NPC, non-polynomial complete), aperiodic self-organization commonly encountered not only in HTSC, but also in protein science [59,60]. From sandpiles to proteins, it appears that the best documented examples of self-organized criticality (SOC) are found in network glasses and HTSC [61].

*Postscript*. Application of "first principle" continuum theories to the FeAs family has uniformly led to the conclusion that electron-phonon interactions cannot cause HTSC in this family [39,40, 62,63], whereas the correct conclusion is that continuum approximations are invalid because of strong disorder and nanoscale phase separation [64-66], as well as the much more subtle consequences of zigzag self-organized percolation (Sec. 8). The latter readily explains how superconductivity in the covalent (non-central interatomic forces) FeAs family can be much more isotropic than in the cuprates [67], yet still retain a lowered dimensionality. Strong disorder explains pseudogap tunneling features in terms of charge density waves [68]. Naturally, percolative conductive paths are phase-sensitive, and when this aspect is combined with the effects of long-range conductive screening of internal electric fields, it is easier to understand why descriptions of disorder using functions dependent on amplitudes alone (such as the participation ratio) are insufficient to describe filamentary self-organization [69].

# Figure Captions

Fig. 1.  The master function for HTSC, $T_c^{max}$ (<R>), provides a least upper bound for bulk layered superconductors, and is believed to be accurate to 10K.  This function is based on the zigzag percolative model for self-organized HTSC dopant networks [28,30].  The original figure [30] has been modified to include the most recent data [34] on BCCOF.  Acronyms as in [30]. Covalent network glasses [20] are centered on <R> = 2.4, but the stabilizing effect of the rigid $CuO_2$ planes shifts the cuprate center to <R> = 2.

Fig. 2. As in Fig. 1, but now the non-cuprate data have been added, so the Figure has become rather "busy".

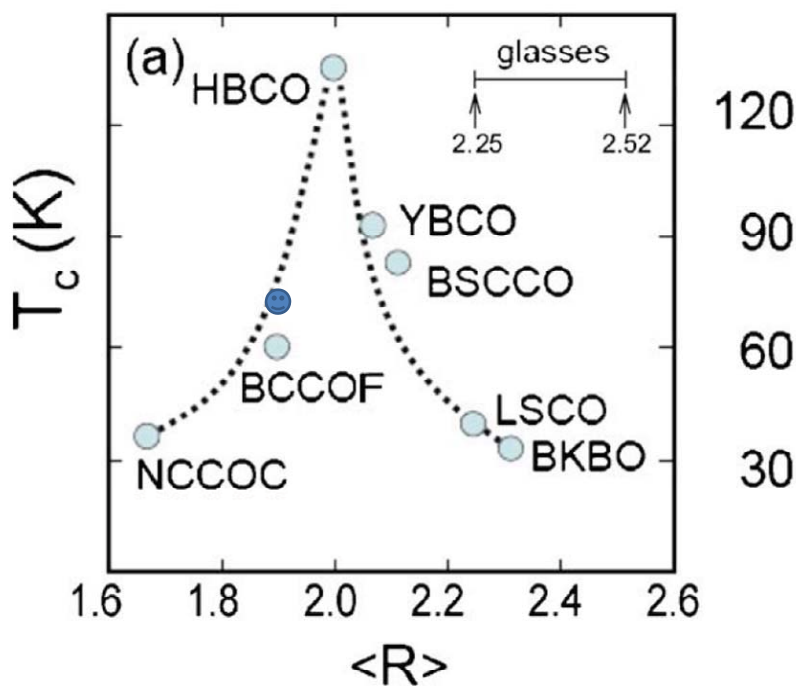

Fig. 1.

21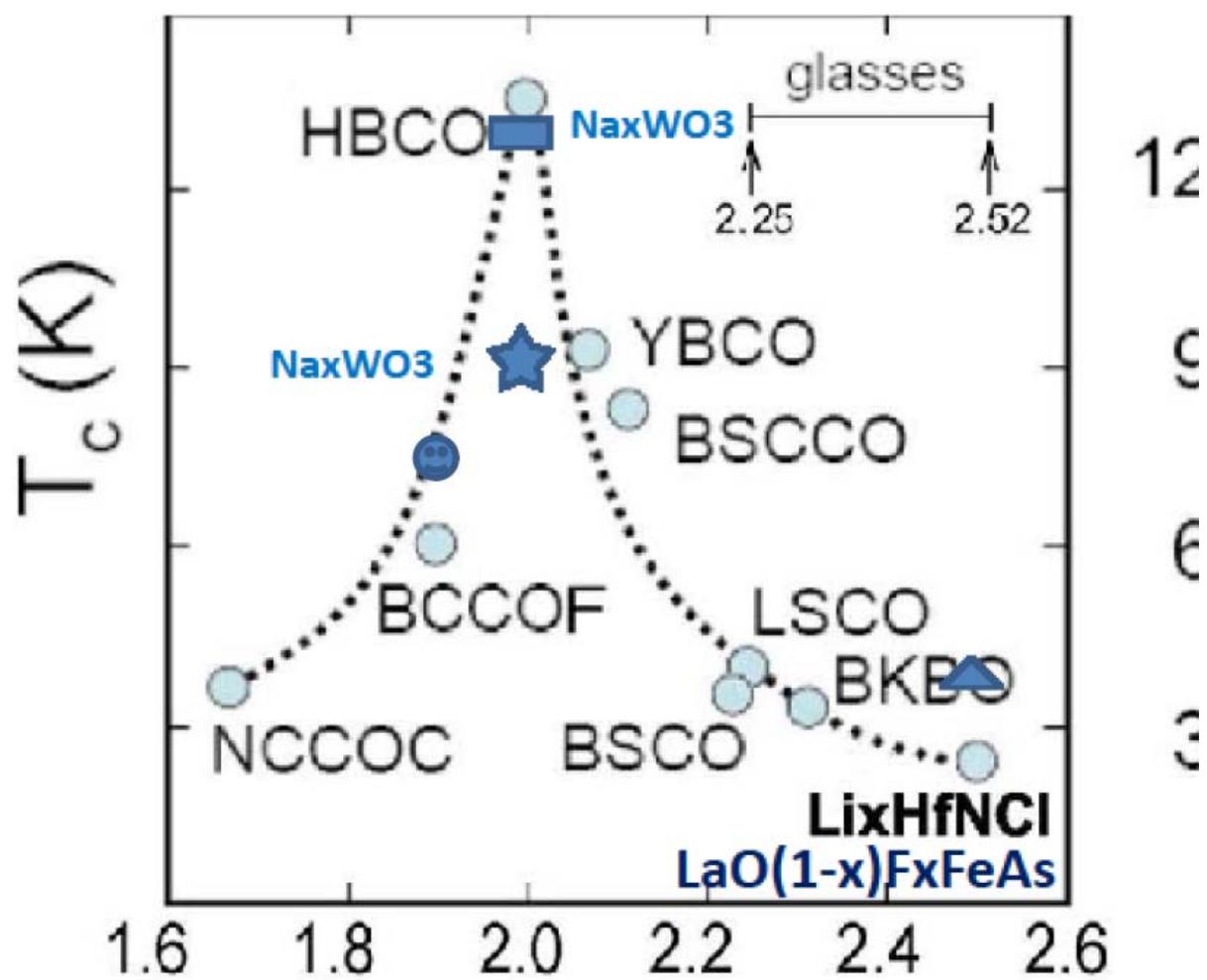

Fig. 2.